\documentclass[
 preprint,
 amsfonts,amsmath,amssymb,
 aps,
 prl,
 ]
{revtex4-1}

\usepackage[utf8]{inputenc}

\usepackage{booktabs}
\usepackage{graphicx}
\usepackage{dcolumn}
\usepackage{bm}
\usepackage{upgreek}
\usepackage{MnSymbol}

\usepackage{xcolor}
\usepackage[colorlinks=true,
            linkcolor={red!50!black},
            urlcolor={green!50!black},
            citecolor={blue!80!black}]{hyperref}

\newcommand{\overbar}[1]{\mkern 1.5mu\overline{\mkern-1.5mu#1\mkern-1.5mu}\mkern 1.5mu}

\def\pd{\partial}

\def\gammabar{\overbar{\gamma}}

\def\varpibar{\overbar{\varpi}}
\def\nubar{\overbar{\nu}}
\def\lambdabar{\overbar{\lambda}}

\def\Pbar{\overbar{P}}

\def\varphibar{\overbar{\varphi}}
\def\nubar{\overbar{\nu}}
\def\lambdabar{\overbar{\lambda}}

\def\abar{\overbar{\beta}}

\def\half{\frac{1}{2}}

\begin{document}
\preprint{QMUL-PH-20-29}

\title{The Forced Soliton Equation and\\ Semiclassical Soliton Form Factors}

\author{Ilarion~V.~Melnikov} \email{melnikix@jmu.edu}
\affiliation{Department of Physics and Astronomy, James Madison University,\\ 901 Carrier Drive, Harrisonburg, VA 22807, USA}
\author{Constantinos Papageorgakis} \email{c.papageorgakis@qmul.ac.uk}
\affiliation{CRST and School of Physics and Astronomy, Queen Mary University of London,\\ Mile End Road, London E1 4NS, UK}
\author{Andrew~B.~Royston} \email{abr84@psu.edu}
\affiliation{Department of Physics, Penn State Fayette, The Eberly Campus,\\ 2201 University Drive, Lemont Furnace, PA 15456, USA}

\date{\today}

\begin{abstract} 
We show that the leading semiclassical behavior of soliton form factors at arbitrary momentum transfer is controlled by solutions to a new wave-like integro-differential equation that describes solitons undergoing acceleration.  We work in the context of two-dimensional linear sigma models with kink solitons for concreteness, but our methods are purely semiclassical and generalizable.
\end{abstract}

\maketitle

\section{Introduction}

Solitons feature prominently in quantum field theories relevant to our current understanding of nature: the Abrikosov and Nielsen--Olesen vortices in the theories of superconductors \cite{Abrikosov:1956sx} and dual strings \cite{Nielsen:1973cs}, baryons in the Skyrme model of nucleonic interactions \cite{Skyrme:1962vh}, and magnetic monopoles in grand unified theories \cite{tHooft:1974kcl,Polyakov:1974ek}.  Furthermore, through semiclassical analysis, solitons provide a direct and appealing connection between solutions to nonlinear classical equations of motion and related quantum field-theoretical observables.

In classical field theory on Minkowski spacetime, topological solitons arise when the field equations admit a set of topologically distinct boundary conditions at spatial infinity, preserving finiteness of the energy. For example, kink solitons exist in two-dimensional models of a single scalar field, $\phi = \phi(t,x)$, with potential $V(\phi)$, when the set of global minima of the potential has multiple components.  The kink is a time-independent energy-minimizing solution to the equations of motion, $\phi = \phi_0(x)$, that interpolates between two distinct minima as $x \to \pm \infty$.  

These disjoint sectors in the space of finite-energy field configurations lead to orthogonal sectors of the Hilbert space of quantum states, and a classical soliton solution corresponds to a one-particle state in a topologically nontrivial sector \cite{Goldstone:1974gf}.  In the two-dimensional scalar field example, the soliton state is fully specified by an on-shell momentum, $P$, and will be denoted $|\Psi_P \rrangle$.  The framework for defining soliton states in quantum field theory, and for carrying out perturbative (semiclassical) computations of observables involving soliton states, was developed in the mid '70s \cite{Dashen:1974ci,Dashen:1974cj,Goldstone:1974gf,Gervais:1974dc,Callan:1975yy,Christ:1975wt,Gervais:1975pa,Tomboulis:1975gf}.  For classic reviews, see \cite{Jackiw:1977yn,Faddeev:1977rm,Rajaraman:1982is}.

Despite the maturity of the subject,  however, seemingly simple yet profound questions remain unanswered.  Consider the case of soliton form factors---matrix elements of field operators between initial and final soliton states, such as: $\llangle\Psi_{P_f} | \widehat{\phi}(t,x) | \Psi_{P_i} \rrangle$. The leading semiclassical behavior of this form factor for generic momentum transfer, $k \equiv P_f - P_i$, is not known in non-integrable models. For momentum transfers much less than the soliton mass, $k \ll M$, the semiclassical form factor is given by the Fourier transform of the classical soliton profile, $\phi_0$ \cite{Goldstone:1974gf}. But, naive attempts to extrapolate this result to momentum transfers $k \sim O(M)$ fail upon comparing to results from integrable models, such as sine--Gordon \cite{Weisz:1977ii}.

In the intervening years, especially after \cite{Seiberg:1994rs},
most work on solitons in non-integrable quantum field theories has centered on quantum-exact results at leading order in the time-derivative, or adiabatic, expansion.  With some notable exceptions \cite{Drukier:1981fq,Bachas:1992dw,Demidov:2015nea}, very little progress has been made in the opposite limit of high-momentum transfer but small coupling.      

The importance---and the underlying difficulty---of the form factor computation is illuminated by considering its role with respect to crossing symmetry of the quantum field theory.  On the one hand, the soliton form factor determines the amplitude, $\mathcal{A}(P_i,k \to P_f)$, for a soliton to absorb a perturbative particle with momentum $k$ created by the field $\phi$:
\begin{align}
 i(2\pi)^2 \delta^{(2)}(k + P_i - P_f) \mathcal{A}(P_i,k \to P_f)  =  \int d^2 x e^{i k_\mu x^\mu} \llangle \Psi_{P_f} | \widehat{\phi}(t,x) |  \Psi_{P_i} \rrangle ~.
\end{align}
On the other hand, by crossing symmetry, this amplitude is equal to the amplitude for a perturbative particle to create a virtual soliton--antisoliton pair:         
\begin{equation}
\mathcal{A}(P_i,k \to P_f) = \mathcal{A}(k \to P_f,-\overline{P}_i)~.
\end{equation}

Hence, if one had access to the soliton form factor at momentum transfers $k \gtrsim O(M)$, and could analytically continue in $k$ to the kinematically allowed region for the pair creation process, then one could determine the pair creation amplitude.  This, in turn, would allow one to quantitatively address  deep and difficult questions in quantum field theory, such as: what is the leading contribution of soliton--antisoliton pairs running in loops to processes involving perturbative particles?  See \cite{Papageorgakis:2014dma} for a recent discussion of this issue, which has received renewed attention in light of various conjectures and computations in certain maximally supersymmetric gravitational and higher-dimensional gauge theories \cite{Douglas:2010iu,Lambert:2010iw,Banks:2012dp,Bossard:2015foa,Bossard:2017kfv}.  
  
In this letter, we report on progress in determining semiclassical soliton form factors at arbitrary momentum transfer.  We follow the phase space path integral formalism of \cite{Gervais:1975pa} and carry out a saddle-point analysis in the one-soliton sector, keeping all time derivatives.  This results in an expansion around solutions to the \emph{forced soliton equation}, a novel second-order wavelike integro-differential  equation for an accelerating soliton.

Although we do not currently know of explicit time-dependent solutions to this equation, we can nevertheless make progress by expanding around a hypothetical solution, by analogy with standard collective coordinate reductions for solitons at small velocity
\cite{Manton:1981mp}.

Using this approach, we construct the \emph{generating functional for semiclassical soliton form factors} depending on a source, $F(t)$.  The generating functional is built from a solution to the forced soliton equation, where the soliton momentum is dictated by Newton's 2nd Law, with the source playing the role of the force: $\dot{P} = F$.  The semiclassical limit of the soliton form factor of any local operator $\widehat{\mathcal{O}}$ is given by  the action of a functional derivative operator, $f_{\mathcal{O}} = f_{\mathcal{O}}[ \frac{\delta}{\delta F}]$, acting on the generating functional. For any local $\mathcal{O}$ we determine the function $f_{\mathcal{O}}$ explicitly in terms of the classical soliton profile.           
   
This result demonstrates that the forced soliton equation is the key to unlocking the semiclassical limit of soliton form factors at arbitrary momentum transfer.  An expanded version of this letter can be found in \cite{Melnikov:2020ret}.

 \section{Saddle-point Approximation in the Soliton Sector}

We work in the class of two-dimensional linear sigma models with classical action
\begin{equation}
S = \int d^2 x \left\{ - \tfrac{1}{2} \partial_\mu \phi \partial^\mu \phi - V_0(m_0;\phi) \right\}~.
\end{equation}
We assume that the minima of $V_0$ are gapped and associated to a spontaneously broken discrete symmetry; the parameter $m_0$ controls the mass gap to the perturbative spectrum.  The coupling $g$ is a parameter in the potential, and we consider potentials with the scaling property
\begin{align}
V_0(m_0;\phi) = \frac{1}{g^2} \widetilde{V}_0(m_0;\widetilde{\phi})\;,  
\end{align}
where $\widetilde{V}_0$ does not depend on $g$ and $\widetilde{\phi} = g \phi$.

In such theories there exists a standard renormalization scheme in which, to all orders in perturbation theory, only the mass parameter is renormalized:
\begin{align}
m_{0}^2 = m^2 + \Delta m^2\;,  
\end{align}
with $m^2$ a finite mass parameter and $\Delta m^2$ the coefficient of the mass counterterm, $V_{\Delta m^2}(\phi)$.  In this scheme the renormalized potential takes the form
\begin{align}
V(\phi) = V_0(m;\phi) + V_{\Delta m^2}(\phi)\;, 
\end{align}
where $V_{\Delta m^2}$ is $O(g^2)$ \cite{Rebhan:1997iv}.  Hence, when we speak of the classical soliton profile, $\phi_0(x)$, we mean a time-independent solution to the equation of motion:
\begin{align}
\partial_{x}^2 \phi - \frac{dV_0}{d\phi}(m;\phi) = 0\;,
\end{align}
An example to keep in mind is $\phi^4$-theory, where
\begin{align}
  \textrm{$\phi^4$ theory:} \quad \Bigg \{\begin{array}{r l} V_0 = & \frac{1}{g^2} \left( g^2 \phi^2 - \tfrac{1}{4} m^2\right)^2 ~, \\ \phi_0 = & \frac{m}{2g} \tanh\left( \frac{m}{\sqrt{2}}(x - X) \right) ~. \end{array}
\end{align}
The constant $X$ represents the kink position.

The renormalized Hamiltonian for any theory in this class takes the form
\begin{equation}
H = \int dx \left\{ \tfrac{1}{2} \pi^2 +  \tfrac{1}{2} (\partial_x \phi)^2 + V(\phi) \right\}~,
\end{equation}
in terms of which the phase-space path-integral representation of matrix elements is
\begin{align}\label{transamp}
\llangle \Psi_f | \widehat{\mathcal{O}} |\Psi_i \rrangle =&~ \int [D\phi D\pi] \Psi_{f}[\phi]^\ast \Psi_{i}[\phi]  e^{i \int dt ( \int dx \pi \dot{\phi} - H)} \mathcal{O}[\pi;\phi] ~,
\end{align}
where $\mathcal{\widehat{O}}$ is any local operator inserted at some spacetime point $(t,x)$, and $\Psi_{i,f}[\phi]$ are initial and final state wavefunctionals.

To define soliton states, we must work in variables appropriate for the soliton sector. This is achieved by a canonical transformation,
\begin{align}
(\pi;\phi) \mapsto (P,\varpi; X,\varphi)\;,  
\end{align}
where the kink position, $X$, has been promoted to a dynamical variable (the collective coordinate), $P$ is its conjugate momentum, and $(\varpi;\varphi)$ encapsulate the remaining field-theoretical degrees of freedom.  The latter are constrained to satisfy
\begin{equation}\label{constraints}
\int dx \;\uppsi_0 \varpi = 0 \qquad \textrm{and}\qquad \int dx\; \uppsi_0 (\varphi - \phi_0) = 0~,
\end{equation}
where
\begin{align}
\uppsi_0 := \frac{1}{\sqrt{M_0}} \partial_x \phi_0  \;,\qquad M_0 := \int dx (\partial_x \phi_{0})^2  
\end{align}
are the normalized zero-mode and classical soliton mass.  Explicitly, the canonical transformation is \cite{Gervais:1975pa,Tomboulis:1975gf,Melnikov:2020ret}
\begin{align}\label{canxfms}
\pi(t,x) =&~  - \frac{(P + \langle \varpi | \varphi' \rangle)}{2 \langle \uppsi_0 | \varphi' \rangle } \uppsi_0(x- X(t)) + \varpi(t,x - X(t)) , \cr
\phi(t,x) =&~ \varphi(t,x-X(t)) ~.  \raisetag{20pt}
\end{align}
Here,
\begin{align}
\langle f | g\rangle := \int d\rho f(t,\rho)^\ast g(t,\rho)\;,  
\end{align}
where $\rho := x - X(t)$ is the co-moving coordinate, and $\varphi'\equiv \pd_{\rho}\varphi$, $\dot \varphi \equiv \pd_{t}\varphi$.

In terms of the new variables, the matrix element \eqref{transamp} takes the form
\begin{align}\label{FFexact}
  \llangle \Psi_{f} | \widehat{\mathcal{O}} | \Psi_{i} \rrangle =&~ \int [DX DP] \int [D\varphi D\varpi D\nu D\lambda]  \Psi_{f}[X,\varphi]^\ast \Psi_{i}[X,\varphi] \times \cr
                                                                   & \qquad\qquad \qquad\qquad\qquad\qquad \times e^{i \int dt ( P \dot{X} + \langle \varpi | \dot{\varphi} \rangle - H_T)} \mathcal{O}[P,\varpi;X,\varphi] ~,
\end{align}
with `total Hamiltonian'
\begin{align}\label{Htot}
  H_{T} =&~ \lambda \langle \uppsi_0 | \varphi - \phi_0 \rangle + \nu \langle \uppsi_0 | \varpi \rangle +  \frac{ (P + \langle \varpi | \varphi' \rangle)^2}{2 \langle \uppsi_0 | \varphi' \rangle^2} +\int d\rho \left\{ \tfrac{1}{2} \varpi^2 + \tfrac{1}{2} \varphi^{\prime 2} + V(\varphi) \right\} ~,
\end{align}
where $\lambda(t),\nu(t)$ are Lagrange multipliers enforcing the constraints \eqref{constraints}.  A soliton state takes the form
\begin{align}
|\Psi_{P}\rrangle = |P \rrangle \otimes |\Psi_{0} \rrangle  
\end{align}
and has wavefunctional
\begin{align}
\Psi_{P}[X, \varphi] = \frac{1}{\sqrt{2\pi}} e^{i P X} \Psi_{0}[\varphi]\;,  
\end{align}
where $\Psi_{0}[\varphi]$ is the groundstate wavefunctional in the $\varphi$-$\varpi$ sector.  The Hamiltonian \eqref{Htot} is nonlocal in space, but local in time, and a diagrammatic Feynman perturbation theory was developed for it in \cite{Gervais:1975pa} under the assumption of small soliton velocity, $P/M_{0} \sim O(g)$.

In this letter our objective is instead to carry out a saddle-point analysis in the $\varphi$-$\varpi$ sector, treating $P(t)$ as an arbitrary background field.  The classical equations of motion in this sector are the constraints
\begin{align}\label{constraints2}
\langle \uppsi_0 | \varphi - \phi_0 \rangle = 0\qquad \textrm{and}\qquad \langle \uppsi_0 |\varpi \rangle = 0\;,  
\end{align}
together with
\begin{align}\label{LMsolve2}
  \dot{\varphi} =&~ \varpi + \nu \uppsi_0 + \beta \varphi' ~,\cr
    \dot{\varpi} =&~ -\lambda \uppsi_0 + \beta\varpi ' + \varphi'' - V_{0}^{(1)}(\varphi) - \uppsi_0' \beta^2 \langle\uppsi_0 |\varphi'\rangle \; . \quad
\end{align}
The $V_{0}^{(1)}(\varphi)$ denotes $\pd V_0(m;\varphi)/\pd\varphi$, and we have introduced the {\it soliton velocity functional}:
\begin{equation}
\beta[\varphi, \varpi; P] := \frac{P + \langle \varpi | \varphi' \rangle}{\langle \uppsi_{0} | \varphi' \rangle^2} ~.
\end{equation}
The moniker is apt since Hamilton's equations in the full theory give $\frac{\partial H_T}{\partial P} = \beta$.

Let $(\varpi,\lambda; \varphi,\nu) = (\varpibar,\lambdabar; \varphibar,\nubar)$ denote a solution to Eqs~\eqref{constraints2}-\eqref{LMsolve2}.  We find that, on a solution, the velocity functional is
\begin{equation}
\beta[\varphibar,\varpibar;P] = \frac{ P + \langle \dot{\varphibar} | \varphibar' \rangle }{\langle \varphibar' | \varphibar' \rangle} =: \abar[\varphibar;P]~,
\end{equation}
and that
\begin{align}\label{backgroundsummary}
& \nubar = -\abar \langle\uppsi_0| \varphibar'\rangle ~, \qquad \lambdabar = \frac{\dot{P}}{\langle \uppsi_0 |\varphibar'\rangle} - \frac{d}{dt} \left( \abar \langle\uppsi_0| \varphibar'\rangle \right) ~, \cr
& \varpibar = \dot{\varphibar} - \abar ( \varphibar' - \langle \uppsi_0 | \varphibar' \rangle \uppsi_0 ) ~,
\end{align}
while $\varphibar$ is a solution to the \emph{forced soliton equation}:
\begin{equation}\label{fse}
\left( \pd_t - \abar[\varphibar;P] \pd_\rho \right)^2 \varphibar  - \varphibar'' + V_{0}^{(1)}(\varphibar) + \frac{\dot{P}(t) \uppsi_0(\rho)}{\langle \uppsi_0 | \varphibar' \rangle} =0 ~,
\end{equation}
additionally satisfying the constraint $\langle \uppsi_0 | \varphibar - \phi_0 \rangle = 0$. The left-hand side of \eqref{fse} takes values in the orthogonal complement of $\mathrm{Span}\{\varphibar'\}$ with respect to the inner-product $\langle~,~\rangle$, and hence the system is not over-constrained.

If one assumes that $P$ is constant, then it is consistent with the equations of motion to also assume that $(\varpibar,\lambdabar; \varphibar,\nubar)$ are time-independent.  The solution to \eqref{fse} is then the boosted soliton profile,
\begin{align}\label{bsp}
\varphibar(\rho) = \phi_0(\gammabar (\rho-\rho_0))\;, 
\end{align}
with the Lorentz factor related to the relativistic momentum as expected \cite{Gervais:1975pa}:
\begin{align}
\gammabar = \sqrt{1 + (P/M_0)^2}\;.  
\end{align}
 The integration constant, $\rho_0$, is fixed by the $\lambda$-constraint ($\rho_0 = 0$ in $\phi^4$-theory).

The assumption of constant $P$ is valid for the transition amplitude---translational invariance guarantees it---but not in the presence of operator insertions: setting $P$ constant leads to results for form factors that are only valid at leading order in a momentum-transfer ($k/M_0$) expansion. 

To continue, we will assume that, for a given $P(t)$, there is a unique solution to the initial value problem associated to the system \eqref{fse} with constraint $\langle \uppsi_0 | \varphibar - \phi_0 \rangle = 0$.  Specifically, under the assumption that $P(t)$ is constant for early enough $t$ with $P(t_i) = P_i$, the initial data will be given by the time-independent solution for $P_i$ just discussed.  The boundary conditions for the soliton as $\rho \to \pm \infty$ will be the standard ones that follow from finiteness of the energy.

Now we expand the Lagrangian, $\langle \varpi |\dot{\varphi} \rangle - H_{T}$, as well as the insertion $\mathcal{O}$, in fluctuations around the solution
\begin{align}
(\varpi,\lambda; \varphi,\nu) = (\varpibar + \delta\varpi,\lambdabar + \delta\lambda; \varphibar + \delta\varphi,\nubar + \delta\nu)\;.  
\end{align}
 The expansion in fluctuations is a $g$-expansion:  one observes that \eqref{backgroundsummary}-\eqref{fse} are consistent with all background fields being $O(g^{-1})$, with order one velocities $\abar$.  

The leading-in-$g$ result for the form factor \eqref{FFexact} reduces to a quantum-mechanical matrix element and is compactly expressed in terms of a semiclassical effective Hamiltonian for the soliton and a semiclassical insertion:
\begin{align}\label{spFF}
& \llangle \Psi_{P_f} | \widehat{\mathcal{O}} | \Psi_{P_i} \rrangle = \frac{1}{2\pi} \int [DXDP] \times e^{i (P_i X_i - P_f X_f)} e^{i \int dt (P \dot{X} - H_{\rm sc}[P])} \mathcal{O}_{\rm sc}[P;X] \times (1 + O(g))~.
\end{align} 
The insertion is given by evaluating $\mathcal{O}$ on the solution,
\begin{align}
\mathcal{O}_{\rm sc}[P;X] := \mathcal{O}[P,\varpibar;X,\varphibar]\;.
\end{align}
 Meanwhile, $H_{\rm sc}$ is determined by a one-loop saddle-point approximation to the \emph{soliton effective Hamiltonian}, $H_{\rm eff}[P]$, defined through
\begin{align}
e^{- i \int dt H_{\rm eff}[P]} :=&~ \int [D\varphi D\varpi D\nu D\lambda] \Psi_{0,f}[\varphi]^\ast \Psi_{0,i}[\varphi] e^{i \int dt ( \langle \varpi | \dot{\varphi} \rangle - H_T)} ~.  \raisetag{20pt}
\end{align}
The effective Hamiltonian admits an expansion in $g$ of the form
\begin{align}
H_{\rm eff} = H_{\rm eff}^{(-2)} + H_{\rm eff}^{(0)} + O(g) \equiv H_{\rm sc} + O(g)\;,
\end{align}
such that $H_{\rm sc}$ captures the tree-level $O(g^{-2})$ and one-loop $O(g^0)$ contributions.  These in turn can be evaluated in terms of the solution, $\varphibar$, to the forced soliton equation.  We find that
\begin{equation}
H_{\rm eff}^{(-2)} = \int d\rho \left\{ \tfrac{1}{2}(1 + \abar^2) \varphibar^{\prime 2} - \tfrac{1}{2} \dot{\varphibar}^2 + V_0(m;\varphibar) \right\}~.
\end{equation}
The Gaussian path integral for the one-loop contribution, $H_{\rm eff}^{(0)}$, is regularized and evaluated in \cite{Melnikov:2020ret}; we will not need the details here.  In the limit of constant $P$, $H_{\rm sc}$ is consistent with the relativistic energy expanded through one loop.

\section{Semiclassical Soliton Form Factors}

Equation \eqref{spFF} may be stated succinctly as
\begin{align}\label{PsiOPsi}
\llangle \Psi_{P_f} | \widehat{\mathcal{O}}[\widehat{\pi},\widehat{\phi}] | \Psi_{P_i} \rrangle = \llangle P_{f} | \widehat{\mathcal{O}}_{\rm sc}[\widehat{P},\widehat{X}] | P_i \rrangle ( 1 + O(g))~,
\end{align}
for a Weyl-ordered operator $\widehat{\mathcal{O}}_{\rm sc}[\widehat{P},\widehat{X}]$.  The dependence of $\mathcal{O}_{\rm sc}$ on the spacetime insertion point, $(t,x)$, occurs in a rather distinct way.  While $t$ appears as the argument of $P(t),X(t)$ in the insertion, $x$ occurs only through the combination $x - X$, as dictated by the canonical transformation \eqref{canxfms} evaluated on the background solution $(\varpi;\varphi) = (\varpibar;\varphibar)$.

We are thus led to consider a general class of quantum-mechanical matrix elements $\llangle P_{f} | \widehat f[\widehat{P},x- \widehat{X}] | P_i \rrangle$ for a Weyl-ordered operator $\widehat{f}$ (and corresponding phase space function $f$).  This motivates the definition of the generating functional,
\begin{align}\label{scsFF}
& \mathcal F_{P_f,P_i}[K,\{F,x\}]  := \frac{1}{2\pi} \int [DX DP] e^{i (P_i X_i - P_f X_f)}  e^{ i \int_{t_i}^{t_f} dt' (P \dot{X} - H_{\rm sc}[P] - P K -(x-X) F) } ~,  \raisetag{20pt}
\end{align}
in terms of which
\begin{align}\label{scformfactor1}
& \llangle P_{f} | \widehat{f}(\widehat{P},x-\widehat{X}) | P_i \rrangle =  \left( f\left[ i \tfrac{\delta}{\delta K(t)}, i \tfrac{\delta}{\delta F(t)} \right] \mathcal F_{P_f,P_i}[K,\{F,x\}] \right) \bigg|_{K,F = 0} ~. \qquad
\end{align}
We refer to $\mathcal F$ as the \emph{generator of semiclassical soliton form factors}.  Since $H_{\rm sc}$ is $X$-independent, the path integral \eqref{scsFF} can be evaluated. The $X$-integral enforces Newton's 2nd Law, via $\delta[\dot{P} - F]$, leading to
\begin{align}\label{scsFFeval1}
\mathcal F_{P_f,P_i}[K,\{F,x\}] =&~ \delta\left( P_f - P_i  - {\textstyle \int_{t_i}^{t_f}} F(t') dt'\right) e^{-i (P_f - P_i)x} e^{ -i \int_{t_i}^{t_f} dt' (H_{\rm sc}[\Pbar] + \Pbar K ) } ~.  
\end{align}
Here, all $F$ dependence is contained in $\Pbar(t)$---obtained as a solution to the 2nd Law---and the single remaining delta function imposes the Impulse-Momentum Theorem.  In the presence of this $\delta$-function,
the solution $\Pbar(t)$ can be written as
\begin{equation}\label{Pbarsol}
\Pbar(t') = \tfrac{1}{2}(P_i + P_f) + \tfrac{1}{2} \left(\int_{t_i}^{t'} - \int_{t'}^{t_f} \right) d\tilde{t} F(\tilde{t})~,
\end{equation}
implying the useful fact
\begin{align}
  \frac{\delta\Pbar(t)}{\delta F(t)} = 0\;.
\end{align}

In order to apply \eqref{scsFFeval1} to evaluate the semiclassical form factor, \eqref{scformfactor1}, we need to investigate the functional derivatives of $\mathcal F$ with respect to $K$ and $F$. The dependence of $\mathcal F$ on $K$ is simple and, thanks to \eqref{Pbarsol}, allows for an explicit evaluation of all $K$ derivatives, resulting in \cite{Melnikov:2020ret}
\begin{align}\label{scformfactor2}
&  \llangle P_{f} | \widehat{f}(\widehat{P},x-\widehat{X}) | P_i \rrangle = \left( f\left[ \tfrac{P_i + P_f}{2}, i \tfrac{\delta}{\delta F(t)} \right] \mathcal F_{P_f,P_i}[0,\{F,x\}] \right) \bigg|_{F = 0} .\quad 
\end{align}
This result is significant: for phase space functions of the form $f_{\mathcal O} = \mathcal{O}_{\rm sc}$, we can use the \emph{constant} $P = \half (P_i + P_f)$ solution for $\varpibar,\varphibar$.  Thus the differential operator $ f_{\mathcal O}\left[ \frac{1}{2}(P_i + P_f), i \frac{\delta}{\delta F(t)} \right]$ appearing on the right-hand side of \eqref{scformfactor2} will be known explicitly, provided the standard soliton solution is known.  

For example, in the case $\widehat{\mathcal{O}} = \widehat{\phi}$, $f_{\mathcal O}$ is given by the boosted soliton profile \eqref{bsp}, with Lorentz factor expressed in terms of the momentum $P = \half (P_i + P_f)$, and spatial argument $\rho = x - X$ replaced with the derivative operator $i \frac{\delta}{\delta F(t)}$.  All nontrivial dependence of the form factor on the momentum transfer is contained in the generating functional $\mathcal{F}_{P_f,P_i}[0,\{F,x\}]$, which depends on the solution to the forced soliton equation with time-dependent $P(t) = \Pbar(t)$, given by \eqref{Pbarsol}.  It can be shown that \eqref{scformfactor2} reproduces known results in the low momentum transfer limit, $|k| \ll M_0$ \cite{Melnikov:2020ret}.

\section{Conclusions}

We have obtained a novel equation---the forced soliton equation.  We have shown that understanding the solutions of the forced soliton equation is the key to understanding the semiclassical behavior of soliton form factors away from the low momentum-transfer limit.  Thus we have, after 40 years, a concrete starting point to address profound questions concerning the nonperturbative structure of quantum field theory.  

A natural first step in studying solutions to \eqref{fse} is to consider perturbation theory in small $\dot{P}$, where the complete diagonalization of the linearized problem around a constant $P$ solution obtained in \cite{Melnikov:2020ret} should be useful.  It would be interesting to see if the framework of \cite{hughes1977well} can shed light on the structure of solutions to the forced soliton equation.  The methods and results presented here are generalizable to other classes of theories with solitons.  An important technical prerequisite is the exact canonical transformation from perturbative to soliton sector phase-space variables.  The latter is known for the magnetic monopole in Yang--Mills--Higgs theory \cite{Tomboulis:1975qt}.


\section*{Acknowledgments}
We acknowledge partial financial support from the NSF PHY-1914505, the Royal Society UF120032 and the STFC ST/P000754/1.

\bibliography{FFs}
\bibliographystyle{apsrev}

\end{document}